# Analyzing Families of Experiments in SE: a Systematic Mapping Study

Adrian Santos, Omar Gómez and Natalia Juristo

**Abstract**—**Context:** Families of experiments (i.e., groups of experiments with the same goal) are on the rise in Software Engineering (SE). Selecting unsuitable aggregation techniques to analyze families may undermine their potential to provide in-depth insights from experiments' results.
**Objectives:** Identifying the techniques used to aggregate experiments' results within families in SE. Raising awareness of the importance of applying suitable aggregation techniques to reach reliable conclusions within families.
**Method:** We conduct a systematic mapping study (SMS) to identify the aggregation techniques used to analyze families of experiments in SE. We outline the advantages and disadvantages of each aggregation technique according to mature experimental disciplines such as medicine and pharmacology. We provide preliminary recommendations to analyze and report families of experiments in view of families' common limitations with regard to joint data analysis.
**Results:** Several aggregation techniques have been used to analyze SE families of experiments, including Narrative synthesis, Aggregated Data (AD), Individual Participant Data (IPD) mega-trial or stratified, and Aggregation of $p$-values. The rationale used to select aggregation techniques is rarely discussed within families. Families of experiments are commonly analyzed with unsuitable aggregation techniques according to the literature of mature experimental disciplines.
**Conclusion:** Data analysis' reporting practices should be improved to increase the reliability and transparency of joint results. AD and IPD stratified appear to be suitable to analyze SE families of experiments.

## 1 INTRODUCTION

In 1999, Basili et al. used the term *family of experiments* to refer to a group of experiments that pursue the same goal and whose results can be combined into joint—and potentially more mature—findings than those that can be achieved in isolated experiments [1]. In particular, families of experiments allow to increase the reliability of the findings [2], increase the statistical power and precision of the results [3], and to assess the impact of experimental changes (i.e., moderators) on results [4], [5], [6].

However, Basili et al.'s definition of family of experiments does not distinguish between two types of groups of experiments: those gathered by means of *systematic literature reviews* (i.e., SLRs, a type of *secondary study* that aims to bring together all the available empirical evidence on a particular topic in a systematic way [7]) and those gathered by means of *experimental replication*—where replications are conducted either by the same researcher or by a group of collaborating researchers that share experimental materials, assist each other during the design, execution, and analysis phases of the experiments, etc. [8], [9], [10]. In our opinion, such groups of experiments should be differentiated as they grant access to different information and, in turn, they may serve to fit different purposes.

For example, while researchers in groups of replications have access to the *raw data* of all the experiments (i.e., a spreadsheet with the response variables and the assignment of subjects to treatments across the experiments), access to the raw data is not guaranteed in SLRs (unless the raw data are requested from the primary studies' authors, and primary studies' authors are willing to share them). Thus, while researchers in groups of replications can apply consistent data-cleaning, data-processing, and data-analysis techniques to ensure that differences across experiments' results are caused only by differences in the data gathered, this is unfeasible in SLRs—as different data-processing and analysis techniques may have been followed to analyze each individual experiment [11]. This may be detrimental to the reliability of SLRs' joint results.

*A. Santos is with the M3S (M-Group), ITEE University of Oulu, P.O. Box 3000, 90014, Oulu, Finland, e-mail: adrian.santos.parrilla@oulu.fi*
*O. Gómez is with Escuela Superior Politécnica de Chimborazo Riobamba, Chimborazo, Ecuador, e-mail: ogomez@espoch.edu.ec*
*N. Juristo is with the Escuela Técnica Superior de Ingenieros Informáticos, Universidad Politécnica de Madrid, Campus Montegancedo, 28660 Boadilla del Monte, Spain, e-mail: natalia@fi.upm.es*

Moreover, while researchers in groups of replications are fully aware of *the experimental settings and the characteristics of the participants* across all the experiments, only the information reported in primary studies is available in SLRs. If this information is scarce or incomplete due to reporting inconsistencies or length restrictions, the appropriateness of SLRs to elicit moderators may be limited (as some moderators may pass unnoticed by the researchers aggregating the results).

Another key difference between groups of experiments built by means of replication and those gathered by means of SLRs is that researchers in the former may opt to introduce *isolated* changes across the experiments with the aim of studying their effects on results. By contrast, experiments gathered by means of SLRs have preset conditions. In turn, if the experiments gathered by means of SLRs differ in more than one element *at a time* (e.g., what may happen when experiments are completely independent—as each experiment may have completely different experimental configurations), differences across experiments' results may come from an "amalgamation" of effects. This may affect the suitability of SLRs to provide moderator effects—as after all, it may be unfeasible to "disentangle" the effects of single moderators on results [12], [13], [14].

Finally, as groups of experiments gathered by means of replication *do not rely on published data* to provide joint results (contrary to groups of experiments in SLRs), they do not suffer from the bias introduced in the results due to selective publication [11]. Thus, groups of replications may provide less-biased conclusions—albeit less generalizable results, as they typically involve fewer experiments—than those provided by SLRs.

To distinguish between groups of experiments gathered by means of SLRs and those gathered by means of replication, we refine Basili et al.'s definition of family of experiments and consider a family to be a group of experiments for which researchers have *first-hand knowledge of all the experiments' settings* and have full access to the *raw data*. In this research, we focus on the techniques that have been used to aggregate experiments' results within families and not on those applied in SLRs—where the *de facto* aggregation technique is meta-analysis of effect sizes [7].

This investigation starts from the observation that minimal research, if any, appears to have been conducted on the aggregation techniques used in SE to analyze families of experiments. The selection of inappropriate aggregation techniques may result in misleading findings and, in turn, to undermining all the effort involved in conducting a family of experiments (e.g., coordinating different research groups to conduct replications across multiple sites, having face-to-face and internet meetings, preparing and translating experimental materials to share among researchers, etc.).

In this article, we conduct a systematic mapping study (i.e., SMS, a type of secondary study where an overview of a specific research area is obtained [15]) with the aim of identifying the techniques that have been used to aggregate experiments' results within families in SE. In addition, we conduct a literature review in mature experimental disciplines, such as medicine and pharmacology, to learn about the advantages and disadvantages of each technique. Finally, we propose a preliminary set of recommendations to *analyze* and *report* families of experiments based on the common limitations found with regard to joint data analysis in SE families. Along our research, we made several **findings**:

- SE families commonly comprise *three* to *five* experiments with *small* and *dissimilar sample sizes*. Families usually involve *different types of subjects* (e.g., professionals and students), and provide *heterogeneous results*.
- From most to least used, Narrative synthesis, Aggregated Data (AD), Individual Participant Data (IPD) mega-trial or stratified, and Aggregation of $p$-values have been used to analyze SE families. Each technique appears to be appropriate in different circumstances, which should be understood before aggregating the experiments' results.
- SE researchers rarely justify the aggregation technique/s used within families. Narrative synthesis and IPD mega-trial are commonly used to analyze families despite their numerous shortcomings according to mature experimental disciplines.
- SE researchers rarely account for heterogeneity of results when providing joint results. SE researchers rarely acknowledge that differences across experiments' results may have emerged due to the natural variation of results and not because of the changes introduced across experiments.

The main **contributions** of this research are a *map and classification* of the techniques used in SE to aggregate experiments' results within families, a *list of advantages and disadvantages* of each aggregation technique according to the literature of mature experimental disciplines, and *a set of recommendations* to analyze and report families of experiments in view of families' common limitations with regard to joint data analysis.

Throughout this article, we argue that it is crucial to understand the advantages and disadvantages of each aggregation technique before applying them, and that the suitability of each technique may be influenced by the characteristics of the family. Blindly applying aggregation techniques without considering their advantages and disadvantages for the specific conditions of the family may result in misleading conclusions and, in turn, to missing a valuable opportunity to extract in-depth insights from experiments' results. Therefore, we make the following suggestions.

---

**Take-away messages**

- The aggregation technique/s used within families should be justified to increase the transparency and reliability of the joint results. Common justifications include the availability of the raw data, the presence of changes across experimental designs or response variable operationalizations, the ability to convey the heterogeneity of the results in intuitive units, the availability of informative plots to summarize the results, and the necessity of interpreting the results in natural units.
- We discourage the use of Narrative synthesis to analyze families since it does not provide a quantitative summary of results and does not take advantage of the raw data to provide joint results or to investigate moderators.
- IPD mega-trial appears to be unsuitable to analyze families of experiments when different types of subjects are evaluated within families (e.g., professionals and students) or when both missing data—due to protocol deviators or drop-outs—and experiments with dissimilar sample sizes are present.
- AD and IPD stratified appear to be suitable to analyze families of experiments. If multiple changes are introduced across experiments within families (as is commonly the case in SE), random-effects models may be more suitable than fixed-effects models.

**Paper organization**. In Section 2, we outline the research method and the research questions of our study. In Sections 3, 4, 5 and 6, we provide answers to each of the research questions. In Section 7, we propose a series of recommendations to analyze families. In Section 8, we present a series of recommendations to report families. We outline the threats to validity of this study in Section 9. Finally, we outline the conclusions of our study in Section 10.

## 2 RESEARCH METHOD

We follow the guidelines proposed by Kitchenham and Charters [16] and those proposed by Petersen et al. [15] for conducting our SMS.

### 2.1 Objectives and Research Questions

The main *objectives* of this study are to systematically identify relevant scientific literature and to map the techniques that have been used to aggregate experiments' results within families from the *viewpoint* of researchers in the *context* of SE. We propose four research questions to meet our objective:

- **RQ1.** How are families of experiments defined and characterized?
- **RQ2.** What techniques have been applied to aggregate experiments' results within families?
- **RQ3.** How do the changes introduced across experiments within families influence the aggregation technique/s used?
- **RQ4.** What limitations with regard to joint data analysis are common across families?

### 2.2 Search and Selection Processes

We iteratively built a search string to identify as many families of experiments as possible. The rationale behind the selection of our final search string is outlined in Appendix A[1]. Our final search string was: *(experiment\*) AND (famil\* OR serie\* OR group\*)*. A total of 1213 documents were retrieved on 14 October 2016 from four databases: IEEE Xplore, ISI Web of Science, Science Direct and Scopus.

We needed to determine when to consider a group of experiments as a family and exercise the decision to define our *exclusion criteria*. We used these exclusion criteria to separate families from other research. However, we faced difficulties during this step. Specifically, we noticed that the distinction between replications and families was not clearly defined in SE. This made us aware that family was an ill-defined term, so we included an additional research question (RQ1) to address this issue. In Section 3, we provide an answer to RQ1 and motivate why we regard three experiments as the lower threshold for considering a group of experiments to be a family.

Eventually, we defined the following *exclusion criteria*:

- The article aggregates fewer than three experiments.
- The article does not compare at least two treatments (e.g., Technology A vs. Technology B) on the same response variable (e.g., quality).

1. The Appendixes can be found in the supplementary material.

- The article does not report experiments conducted with human participants.
- The article is duplicated.
- The article is not peer-reviewed (e.g., it is a call for papers, keynote speech, preface)

We dismissed an article whenever at least one of the exclusion criteria was met. Table 1 provides a summary of the selection process that we undertook to select families of experiments. In particular, in Stage 0, the first author went through the list of articles—sorting them by title, year and author/s—to eliminate duplicates and non-relevant documents. The number of articles remaining at the end of this stage was 572.

In Stages 1, 2 and 3, the first two authors excluded articles by title, abstract reading and in-depth reading, respectively. In case of disagreement, the third author helped to make the decision of whether the article would proceed to the next stage. Table 1 shows the number of disagreements and the number of articles proceeding to the next stage. At the end of Stage 3, 36 articles were considered to be relevant. These 36 articles constituted our initial set of primary studies, which we used during Stage 4 to perform a *backward snowballing process* (i.e., a procedure where relevant studies are gathered from the primary studies' reference lists [17]). During Stage 4, we identified three new articles. These articles were not indexed in the previous search because (1) they referred to a group of experiments as a "set"; (2) they contained the "meta-analysis" keyword but no term referring to the set of experiments; and (3) they used the term "replication" to refer to a family of experiments. After all the stages were complete, we identified a total of 39 primary studies.

### 2.3 Extraction Process

We designed a data extraction form (Appendix B) to extract all the relevant data from each primary study. We extracted a total of 13 fields of information. We improved each field of information after a first round of screening made by the first author. The purpose of this first screening was to establish categories to classify the aggregation techniques used in families and the different characteristics that may have influenced the selection of such techniques.

The first and second authors gathered the information of each primary study independently using the final data extraction form. 12 out of 39 articles contained at least one field where a conflict materialized. The first and second authors consulted to the third author to reach a final agreement in those cases. The conflicts were related mostly to the experiments' sample sizes (some articles reported the final number of subjects after data pre-processing and not others) and the dimensions changed across the experiments (e.g., in cases where it was not clear whether response variables or protocols were changed across experiments). For example, to solve the inconsistencies in the case of sample sizes, we made the decision of using the final number of subjects after pre-processing. In the case of doubts about whether certain changes had been introduced across experiments within families, we made a guess based on the aggregation technique/s used.

TABLE 1
Summary of results across stages.

| Stage | Goal | Total | Excluded | Disagree | Included |
|---|---|---|---|---|---|
| Stage 0 | Remove duplicates | 1213 | 633 | - | 580 |
| Stage 0 | Remove non-relevant documents | 580 | 7 | - | 572 |
| Stage 1 | Exclude studies by title | 572 | 383 | 23 | 189 |
| Stage 2 | Exclude studies by abstract | 189 | 139 | 15 | 50 |
| Stage 3 | In-depth reading | 50 | 14 | 7 | 36 |
| Stage 4 | Snowballing process | 36 | - | - | 3 |

## 3 RQ1: Family Definition and Attributes

We started this study following Basili et al.'s definition of family of experiments [1]: a group of experiments that pursue the same goal to extract mature conclusions. However, we soon realized that this definition did not provide a clear cut-off to distinguish between groups of replications and groups of experiments gathered by means of SLRs. In particular, while the scope of an SLR is likely to be wide (as the goal is to combine all available research on a particular topic into a joint result), in groups of replications, the goal tends to be narrower (since a small set of hypotheses on a limited set of response variables is usually assessed). Moreover, in SLRs, only the information reported in the primary studies is known. Thus, some relevant information (e.g., characteristics of the participants or experimental settings) may pass unnoticed if not fully reported due to length restrictions or reporting inconsistencies—unless the experimenters are contacted to share more detailed information or even to share the raw data. Conversely, in groups of replications—conducted by either the same researcher or a group of collaborating researchers from the same or different groups and/or institutions—access to more detailed information about the experiments is guaranteed. For example, in groups of replications, researchers typically share laboratory packages and instructional materials to ease the execution of experiments [18], assist each other via in-person or internet meetings during the experiment's planning or design phases, assist each other during the execution of replications [19], etc. In turn, this close collaboration leads to greater knowledge of all the experiments' settings and guarantees full access to the raw data of all the experiments. This may increase the reliability of the joint results and, in cases where experiments' results differ, may ease the elicitation of moderators. Thus, for us, a major difference between groups of experiments gathered by SLRs and those gathered by means of coordinated replications is first-hand knowledge of all the experiments' settings and access to the raw data.

As a consequence of the search that we conducted, we also realized that there was no exact cut-off point to discern between (1) a series of planned, coordinated, or opportunistic replications conducted by a sole researcher—or group of collaborating researchers—and; (2) isolated replications conducted by researchers that do not interact or collaborate with those who run the baseline experiments.

To begin answering RQ1, we consider a group of replications to be a family if:

- **At least two treatments** (e.g., Technology A vs. Technology B) are explicitly exercised and compared within all the experiments on **a common response variable** (e.g., quality). Three or more treatments may be compared, but two is the minimum. Studies where a treatment is compared with the results reported in the literature (e.g., reported industry averages) are excluded. This way we ensure that the statistical assumptions required by some aggregation techniques (e.g., normality or equality of variances [20]) can be thoroughly checked before providing joint results.
- **At least three experiments** are included within the family, so it is possible to provide joint results and to study *experiment-level* moderators (e.g., programming language, testing tool). In particular, if the goal is to assess experiment-level moderators within families with less than three experiments, some techniques, such as meta-regression (i.e., one of the procedures for studying moderators with meta-analysis of effect sizes [21]) cannot be applied, as meta-regression requires a minimum of at least three experiments—because a regression line fitted to just two data points would explain all the variability in the data. By using this lower bound of three experiments (1) we avoid the problem of considering an isolated replication of a baseline experiment as a family of experiments and (2) we ensure that all aggregation techniques can be applied within families, irrespective of whether the goal is to provide joint results or to investigate moderators.
- **Full access to all experiments' raw data** is guaranteed to ensure that homogeneous data processing, cleaning and analysis steps can be applied to each experiment before providing a joint result. Thus, sets of replications gathered from the literature are omitted from the definition of family that we propose here.
- **First-hand knowledge of the settings** is guaranteed in all the experiments, so it is possible to minimize the impact of potentially unknown factors on the results and, in cases where the results differ, to hypothesize on likely moderators behind the differences of results.
- **Different subjects participate in each experiment** within the family so the data are independent across the experiments. As otherwise, if the same subjects participate across the experiments, subjects' scores may be correlated across the experiments, and if this correlation is not correctly accounted for when providing joint results, joint results may be biased [22].

TABLE 2
Families of experiments' characteristics (ordered by family size).

| Size | Sample Sizes | Type of subjects | Raw-data | Package | Venue | Year | ID |
|---|---|---|---|---|---|---|---|
| 12 | 215 (16, 16, 16, 24, 22, 22, 18, 24, 16, 16, 15, 10) | Students and professionals | ✓ | ✓ | PROFES | 2015 | [P1] |
| 8 | 455 (42, 39, 29, 35, 31, 31, 172, 76) | Undergraduates | ✗ | ✓ | ICST | 2012 | [P2] |
| 6 | 126 (29, 44, 53, ?, ?, ?,) | Students and professionals | ✗ | ✓ | TSE | 2016 | [P3] |
| 5 | 177 (20, 15, 29, 87, 26) | Students and professionals | ✗ | ✗ | JSS | 2005 | [P4] |
| 5 | 232 (72, 28, 38, 23, 71) | Students, type unknown | ✗ | ✗ | MODELS | 2008 | [P5] |
| 5 | 284 (55, 178, 13, 14, 24) | Students and professionals | ✗ | ✓ | EMSE | 2009 | [P6] |
| 5 | 80 (6, 13, 16, 13, 32) | Graduate and undergraduates | ✗ | ✗ | AOSD | 2011 | [P7] |
| 5 | 112 (24, 24, 28, 20, 16) | Students and professionals | Partially | ✓ | TSE | 2013 | [P8] |
| 5 | 594 (48, 214, 118, 137, 77) | Undergraduates | ✗ | ✗ | EMSE | 2014 | [P9] |
| 5 | 74 (10, 22, 16, 13, 13 ) | Graduate and undergraduates | ✗ | ✗ | EMSE | 2014 | [P10] |
| 5 | 55 (7, 22, 6, 9, 11) | Graduate and undergraduates | ✗ | ✗ | TOSEM | 2015 | [P11] |
| 4 | 72 (44, 15, 9, 4) | Students and professionals | ✗ | ✗ | JSS | 2006 | [P12] |
| 4 | 94 (31, 25, 18, 20) | Students and professionals | ✗ | ✗ | IST | 2010 | [P13] |
| 4 | 74 (13, 35, 18, 8) | Graduate and undergraduates | ✓ | ✗ | TSE | 2010 | [P14] |
| 4 | 111 (48, 25, 19, 19) | Graduate and undergraduates | ✗ | ✗ | TSE | 2011 | [P15] |
| 4 | 139 (33, 51, 24, 31) | Graduate and undergraduates | ✗ | ✓ | TOSEM | 2014 | [P16] |
| 4 | 86 (24, 22, 22, 18) | Students and professionals | ✓ | ✓ | TOSEM | 2014 | [P17] |
| 4 | 92 (28, 16, 36, 12) | Graduate and undergraduates | ✗ | ✓ | IST | 2015 | [P18] |
| 4 | 88 (25, 25, 23, 15) | Students and professionals | ✓ | ✓ | TOSEM | 2015 | [P19] |
| 4 | 81 (11, 16, 22, 32) | Graduate and undergraduates | ✗ | ✓ | EMSE | 2016 | [P20] |
| 3 | 66 (24, 24, 18) | Students and professionals | ✗ | ✗ | EMSE | 1998 | [P21] |
| 3 | 60 (20, 20, 20) | Professionals | ✗ | ✗ | TSE | 2001 | [P22] |
| 3 | 24 (8,8,8) | Professionals | ✗ | ✗ | IST | 2004 | [P23] |
| 3 | 34 (9, 12, 13) | Graduate and undergraduates | ✗ | ✗ | IST | 2004 | [P24] |
| 3 | 115 (60, 26, 29) | Graduate and undergraduates | ✗ | ✗ | ISMS | 2005 | [P25] |
| 3 | 34 (14, 8, 12) | Graduate and undergraduates | ✗ | ✗ | ICSE | 2008 | [P26] |
| 3 | 15 (8, 5, 2) | Graduates | ✗ | ✗ | EMSE | 2009 | [P27] |
| 3 | 143 (78, 29, 36) | Undergraduates | ✗ | ✗ | IST | 2011 | [P28] |
| 3 | 172 (53, 98, 21) | Professionals | ✗ | ✗ | IST | 2011 | [P29] |
| 3 | 84 (30, 45, 9) | Graduate and undergraduates | ✗ | ✗ | IST | 2012 | [P30] |
| 3 | 75 (33, 18, 24) | Graduate and undergraduates | ✗ | ✗ | RESER | 2012 | [P31] |
| 3 | 215 (70, 73, 72) | Undergraduates | ✗ | ✗ | EMSE | 2012 | [P32] |
| 3 | 79 (19, 31, 29) | Undergraduates | ✗ | ✗ | IST | 2013 | [P33] |
| 3 | 45 (14, 12, 19) | Graduate and undergraduates | ✗ | ✓ | SEKE | 2013 | [P34] |
| 3 | 64 (12, 32, 20) | Graduates | ✓ | ✗ | JSS | 2013 | [P35] |
| 3 | 75 (33, 18, 24) | Undergraduates | ✗ | ✓ | EMSE | 2014 | [P36] |
| 3 | 92 (20, 25, 47 ) | Graduates | ✗ | ✗ | QRS | 2014 | [P37] |
| 3 | 91 (35, 22, 34) | Undergraduates | ✗ | ✗ | IST | 2015 | [P38] |
| 3 | 169 (40, 51, 78) | Graduate and undergraduates | ✗ | ✓ | IST | 2015 | [P39] |

Table 2 shows the 39 families of experiments that we identified. Families are ordered in Table 2 according to family size (i.e., the number of experiments). The columns of Table 2 show the sample size of the experiment, the types of subjects participating in the family, whether the raw data and laboratory package were provided, the publishing venue of the family, the publication date and the reference.[2]

With regard to the number of experiments within families (first column), 48% of the families include three experiments, 23% include four experiments, and 20% include five experiments. Three larger families comprise six, eight and 12 experiments. The average number of experiments included within families has increased over time, which suggests an improvement in the maturity of the area (since the larger the number of experiments within families, the larger the sample size, and thus, the larger the potential reliability of the joint results). However, the increase was slight: approximately three experiments were included within families between 1998 and 2006, while the average number of experiments in the most recent years increased to four and peaked at five in 2015.

In terms of the number of subjects within families (second column), 12% of the 39 families contained fewer than 50 subjects and 48% contained between 50 and 100 subjects. The larger the number of subjects, the smaller the number of families. A total of 12% of the families included more than 100 subjects but fewer than 150. With regard to sample size (second column in parentheses), most of the experiments identified within families include 10 to 30 subjects. The number of experiments containing between 1 and 9 subjects is roughly equal to the number of experiments with 30 to 39 subjects (16 experiments and 19 experiments, respectively). As recent families have, on average, five experiments and 25 subjects participating in each experiment, an increase in the total number of subjects is observed in recent years.

Regarding the type of subjects involved within families (third column), 38% of the families include both graduate and undergraduate students, 25% include both students and professionals, and 18% include only undergraduates. A total of 66% of the families include only students. Only 7% of the families include just professionals. The total number of professionals participating within families over time does not appear to follow any pattern: the number remains constant at approximately 20 per year.

With regard to raw data (fourth column) and laboratory package (fifth column), the raw data were provided in 15%

---

2. In Appendix C, we provide a series of figures to visualize the data shown in Table 2.

of the families (although only 7.5% were accessible as of March 2018) and the laboratory package was provided in 43% of the families (although most were not accessible as of March 2018). While experimental packages appear to be provided more often from 2012 onwards, raw data provision does not appear to have increased over time. This situation prevents re-analysis by third-party researchers with perhaps more appropriate aggregation techniques than those applied in the original articles.

In terms of publishing venue (sixth column), IST published 25% (10 of 39) of the families, EMSE published 20% (8 of 39), TSE published 13% (5 of 39) and TOSEM published 10% (4 of 39). The remaining primary studies were published in other venues.

In terms of publication date (seventh column), a steady number of families is observed between 1998 and 2004, whereas a sharp increase is observed from 2008 onwards. This increase is possibly the result of the growing interest in experimentation and the recent calls for replication in SE. Again, the field of SE experimentation might be maturing.

Finally, we want to make one last observation. In mature experimental disciplines such as medicine, the closest representative of families of experiments (i.e., multicenter clinical trials [20]) are run with pre-established protocols defining the experimental settings and the set of procedures that must be strictly adhered to during the execution and analysis of the experiments [23], [24], [25]. Moreover, in multicenter clinical trials aiming to assess the efficacy of new drugs, the populations under assessment across all the centers are specifically defined to ensure consistency of the results and to avoid confounding effects [20], [26], [24]. By contrast, most SE families are formed without any a priori plan, and changes are commonly introduced across experiments opportunistically, either to increase the generalizability of results or to assess moderators. Furthermore, analysis decisions within SE families are commonly driven by the results of statistical tests (e.g., tests of normality [27]), or examples from other researchers and personal preferences (e.g., when authors conduct analyses similar to those undertaken in previous families). This characteristic conforms to the findings reported in medicine years ago [28].

As a summary, SE families tend to have the following characteristics:

- Most families comprise **three to five experiments** with dissimilar and small sample sizes (i.e., fewer than 30 subjects per experiment) with a total of **approximately 100 participants**.
- Most families **include only students**. However, different types of students are commonly involved. Professionals are the only participants in just three out of 39 families.
- Almost no family provided the raw data. Less than half of the families provided a laboratory package. **Almost none of them are currently accessible**.
- **Families appear to be "happenstance"**. In other words, families do not follow a pre-specified protocol outlining the procedures for either conducting experiments, analyzing the experiments, or aggregating the results.

- Most families are published in **journals (IST, EMSE, TSE, or TOSEM)**.

## 4 RQ2: ANALYSIS TECHNIQUES APPLIED

Table 3 shows the techniques that have been used to aggregate experiments' results within families (from most to least used), the number of families that apply each technique, and the references of the families.[3]

TABLE 3
Aggregation technique by family of experiments.

| Technique | N | Primary Studies |
|---|---|---|
| Narrative synthesis | 18 | [P21][P25][P4][P12][P14][P13] [P29][P2][P32][P35][P37][P10] [P16][P9][P36][P11][P19][P20] |
| AD | 15 | [P22][P24][P5][P6][P28][P30] [P35][P8][P33][P17][P39][P18] [P1][P38][P20] |
| IPD mega-trial | 13 | [P23][P26][P27][P14][P28][P15] [P7][P30][P34][P33][P10][P39] [P11] |
| IPD stratified | 6 | [P21][P26][P31][P16][P36][P3] |
| Aggregation of $p$-values | 3 | [P22][P24][P29] |

**Narrative synthesis** was used to analyze 46% of the families. Narrative synthesis is an aggregation technique that provides a *textual* summary of results as a joint conclusion [29]. Families applying Narrative synthesis do not explicitly mention any term to refer to this aggregation technique (although some use the term "global analysis" [P13], [P4]). In general, it is difficult to distinguish whether authors are aggregating the experiments' results or just comparing them. For example, Scanniello et al. [P1] summarizes results as *"...this is true of all the experiments, the only exception being [experiment X] on the [response variable] GD, where the statistical test returned a p-value equal to 0.39..."*, and Staron et al. [P12] summarized the results as *"...in general the stereotypes improve... and half of the results were statistically significant..."*. The main advantage of Narrative synthesis is that, as only a textual summary of the results is provided, Narrative synthesis enables the combination of the findings of experiments with wildly different experimental designs, response variables or statistical tests into joint results [30], [29]. Moreover, in Narrative synthesis, discordance across experiments' results is not seen as a threat to the validity of the joint result but as an opportunity to study the effect of moderators [6]. For example, in cases where results differ across experiments, Ali et al. [P37] claim *"...one plausible explanation is that participants had more experience with standard UML state machines before the experiment than Aspect state machines and hence their understandability..."*, and Ricca et al. [P26] claim *"...the use of stereotypes does not always introduce significant benefits... the provided material is different, and this can be the reason for different results..."*.

Although Narrative synthesis is straightforward to use, it has some relevant shortcomings. Specifically, Narrative synthesis does not provide a joint effect size or $p$-value, which hinders the incorporation of results in prospective studies. Another major shortcoming is that a subjective

---
3. We map primary studies to analysis technique/s in Section 5.

weight—dependent upon the analyst—is assigned to each experiment towards the overall conclusion [21] (e.g., should all experiments be weighted identically regardless of their sample sizes?; should experiments with professionals be weighted more than those with students since professionals are more representative of reality?). In turn, Narrative synthesis [21] may hinder the reliability and reproducibility of the results.

**Aggregated Data** meta-analysis (i.e., AD) was used in 38% of the families. AD is commonly known in SE as *meta-analysis of effect sizes* and is the *de facto* aggregation technique to aggregate experiments' results in SLRs [7]. Families' authors refer to AD simply as *meta-analysis*: "the set of statistical techniques used to combine the different effect sizes of the experiments" [P8]. Effect sizes quantify the relationship between two groups (or more generally, between two variables: the dependent and the independent variables [21]). Effect sizes can be computed from experiments' summary statistics (e.g., means, standard deviations, and sample sizes [31]) or from statistical tests' results (e.g., the $t$-test's $t$-value and degrees of freedom [27]).

Effect sizes are commonly divided into two main families [32]: the r family and the d family. The r family's effect sizes quantify the strength of the relationship between two variables. The strength of this relationship is usually presented as a Pearson correlation [27]. For example, Gonzalez et al. [P18] calculated the Pearson correlation between two variables in each experiment and aggregated the results using AD. Manso et al. [P5] followed an identical procedure but used the Spearman correlation instead (a non-parametric correlation coefficient [27]). The d family's effect sizes quantify the difference between the means of two groups. The size of the difference is usually conveyed in *standardized* units to rule out differences across experiments' response variable scales. Cohen's d or Hedges' g [33], [34] are common representatives of the d family in SE [35]. For example, Fernandez et al. [P35] calculated the Hedges' g of each experiment and then aggregated all of them into a joint result by means of AD.

AD is a statistical technique that delivers a *weighted* average of all experiments' effect sizes as a joint effect size [21]. In general, the *weight* given to each experiment is directly proportional to the sample size—if a fixed-effects model is used [21]—or to the sample size of the experiment and the total *heterogeneity* of the results (i.e., the variability of results that cannot be explained by natural variation)—if a random-effects model is used [21].[4]

One of the main advantages of AD is that it can be used to combine the results of experiments with different designs and response variable scales into a joint conclusion—as long as a suitable standardized effect size can be calculated [21], [36]. In addition, AD can easily handle the heterogeneity of results (by simply fitting a random-effects model instead of a fixed-effects model [21]) or elicit experiment-level moderators (e.g., by using meta-regression or subgroup meta-analysis [21]). Other advantages of AD are its intuitive visualizations (i.e., forest plots [21]) and its straightforward statistics to quantify heterogeneity (e.g., $I^2$ [21]).

---
4. Assuming a common variance across experiments.

Figure 1 shows an example of a forest plot. As shown in Figure 1, the effect size of each experiment is represented by a square. The size of each square represents the weight of the effect size in the overall result (i.e., the black diamond at the bottom), and the width of the line crossing each square represents the uncertainty of the effect size in each experiment (i.e., its 95% confidence interval [36]). The assessment of the heterogeneity is also straightforward by means of the $I^2$ statistic (e.g., the heterogeneity is considered to be small, medium or large if the $I^2$ statistic is larger than 25%, 50% or 75% respectively [21]).

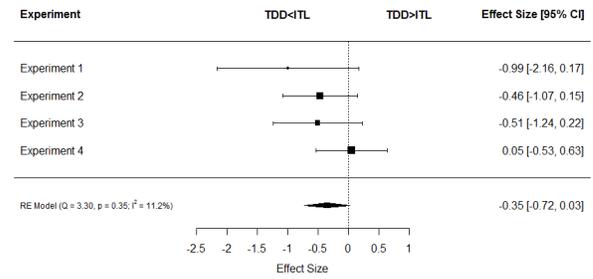

Fig. 1. Forest plot: AD example.

Though appealing, one of the main limitations of AD is that it cannot simultaneously assess the effects of multiple factors on the results (e.g., the effects of the treatments, experimental tasks and their interactions). Instead, AD is commonly applied to aggregate experiments with relatively simple designs [21]. Another shortcoming of AD is that the the effect sizes' statistical assumptions need to be checked before providing joint results (e.g., normality or homogeneity of variance for Cohen's d [37][38][39]).

**Individual Participant Data mega-trial**[5] (i.e., IPD mega-trial) was used to analyze 33% of the families. In IPD mega-trial, the raw data of all the experiments are *pooled* together into a joint dataset and then analyzed as if the raw data were obtained from a single "big" experiment [P27]. Families' authors name IPD mega-trial after the statistical model applied (e.g., *ANOVA* and *GLM* [P11]). As an example of its application, Cruz et al. [P28] individually analyzed each experiment by means of the Kruskal-Wallis test (i.e., the non-parametric counterpart of the one-way ANOVA [27]) and then pooled all the experiments' raw data into a joint dataset to analyze the data jointly with the same test. Ricca et al. [P26] followed an identical procedure but used the Wilcoxon test (i.e., the non-parametric counterpart of the dependent $t$-test [27]). Table 4 shows the IPD mega-trial statistical models that were used within families. As Table 4 shows, non-parametric tests are dominant.

Despite its intuitiveness, IPD mega-trial may provide biased results if the experiments are unbalanced across treatments (e.g., due to missing data caused by protocol deviators or drop-outs) and have different sample sizes, or if the subjects within the same experiment resemble more to each other than to those of other experiments [41],

---
5. The term "mega-trial" is not to be mistaken with the term "multi-center" trial. While mega-trial refers to an analysis approach [28], "multicenter" trial refers, in medicine, to multiple experiments conducted at different sites with a common underlying protocol [20], [40].

TABLE 4
IPD mega-trial statistical model by family.

| Statistical test | Primary studies |
|---|---|
| Non-parametric | [P26][P27][P28][P15][P33][P39][P11][P10] |
| ANOVA | [P14][P7][P30] |
| Others | [P23][P34] |

[42], [43] (e.g., what may occur when experiments with either professionals or students are run). As a result, the use of IPD mega-trial is commonly discouraged in mature experimental disciplines [41], [42], [43].

Another disadvantage of IPD mega-trial is that some statistical tests cannot jointly analyze experiments with different designs. For example, if within-subjects experiments (i.e., repeated-measures experiments) and between-subjects experiments are tried to be analyzed together by means of a repeated-measures ANOVA [27], the repeated-measures ANOVA "throws away" all the data coming from the between-subjects experiments (because data in such experiments are not repeated within subjects). Additional shortcomings of IPD mega-trial are that it requires all the experiments to use identical response variable scales—as otherwise, the joint effect provided may be biased [11]—and that the statistical models are built on top of some statistical assumptions that need to be checked before interpreting the results (e.g., normality or homogeneity of variance [27]). Finally, the heterogeneity of results across experiments cannot be included within IPD mega-trial statistical models, as the experiment from which the raw data are from is not accounted for in IPD mega-trial. However, IPD mega-trial's statistical flexibility is an advantage: some statistical models (e.g., ANOVA [27]) allow the inclusion of as many factors as desired to model the relationship between the data and the characteristics of the family. For example, Ricca et al. [P14] fitted an IPD mega-trial ANOVA model with experience (i.e., undergraduate, graduate, research assistant) and separately fitted another ANOVA with ability (high and low) to assess the effect of experience and ability on the results.

**Individual Participant Data stratified** (i.e., IPD stratified) was applied in 15% of the families. As in IPD mega-trial, IPD stratified involves the central collection and processing of all the experiments' raw data into a joint dataset. However, instead of analyzing the raw data jointly as coming from a single "big" experiment, in IPD stratified, a factor representing the experiment that is the source of the raw data is included in the statistical test [28]. This relationship is considered within statistical models by including an extra factor (i.e., "Experiment") within the statistical models fitted. As an example, commonly applied IPD stratified statistical tests are ANOVA models accounting for two factors: "Treatment" and "Experiment". The authors of families of experiments using IPD stratified refer to it with the name of the technique applied (e.g., *ANOVA* [P21] or *linear regression* [P3]), although some have used the name *comprehensive analysis* [P16]. As an example of its application, Runeson et al. [P31] individually analyzed each experiment with a Wilcoxon test and then analyzed the family as a whole with an ANOVA model including "Treatment" and "Experiment" as factors. Table 5 shows the IPD stratified statistical models applied within families. Table 5 shows that ANOVA is the most commonly used technique in IPD stratified [27].

TABLE 5
IPD stratified statistical model by family.

| Statistical test | Primary studies |
|---|---|
| ANOVA | [P21][P26][P31][P36] |
| Linear regression | [P3] |
| Permutation test | [P16] |

Contrary to IPD mega-trial, IPD stratified allows the inclusion of the heterogeneity of results across experiments. In addition, IPD stratified enables the inclusion of flexible statistical assumptions within statistical models (e.g., different variances across experiments vs. identical variance across experiments) to increase the reliability of the joint results [20]. IPD stratified is considered to be the *gold standard* in medicine for analyzing multicenter clinical trials [44], [45], [28].

The main shortcomings of IPD stratified are its complexity for assessing the heterogeneity of results (as no straightforward statistic such as $I^2$ is produced by IPD stratified models), its reliance on identical response variables across experiments for providing joint results, and the difficulty of fitting and checking the statistical assumptions of some relatively complicated statistical models (e.g., Linear Mixed Models [20]).

Finally, **Aggregation of** $p$**-values** was used in just 7% of the families. In Aggregation of $p$-values, the *one-sided* $p$-values of all the experiments are pooled by means of a statistical method such as Fisher's or Stouffer's [21].[6] For example, Laitenberger et al. [P22] analyzed each experiment with a *one-sided* dependent $t$-test and then pooled the $p$-values into a joint result by means of the Fisher method. The main advantage of Aggregation of $p$-values is that it can aggregate the $p$-values of experiments with any design, response variable or statistical test into joint results [46], [47], [20]. The main disadvantage of Aggregation of $p$-values is that it cannot provide a joint effect size—but only a joint $p$-value. In turn, as $p$-values confound effect size and sample size (i.e., a small $p$-value may emerge because of a relevant effect size or due to a huge sample size and an almost negligible effect size [36]), the interpretability of the results is not straightforward [21], [20]. Another major disadvantage of Aggregation of $p$-values is that in its basic procedures (e.g., Fisher's or Stouffer's methods [21]) an identical weight is assigned to each experiment, regardless of its quality or sample size. This identical weight assignment may affect the reliability of the joint results [20].

Table 6 shows a summary of the advantages and disadvantages of each aggregation technique according to mature experimental disciplines.

## 5 RQ3: TECHNIQUE SELECTION WITHIN FAMILIES

The changes introduced across the experiments within families may impact the suitability of the aggregation tech-nique/s applied. For example, Runeson et al. [P36] claimed that as many changes were made in the third experiment, the results could not be aggregated [P36]. This argument was also used previously in SE to prevent the aggregation of experiments' results [48]. Thus, we believed that it would be sensible to investigate whether making certain changes across the experiments within families hindered or benefited the application of certain aggregation techniques.

---
6. Although more advanced Aggregation of $p$-values techniques also exist [20], only Fisher's and Stouffer's methods [21] have been used in SE.

TABLE 6
Advantages and disadvantages of the analysis techniques.

| Advantages | Technique | Disadvantages |
|---|---|---|
| ✓ Fast interpretation of results<br>✓ Intuitive approach<br>✓ Independent of design, metric and statistical test | **Narrative synthesis** | ✗ **Provides no effect size nor $p$-value**<br>✗ Subjective weighting<br>✗ Non-reproducible results |
| ✓ **Independent of design and metric**<br>✓ Straightforward visualizations<br>✓ Moderators and heterogeneity | **AD** | ✗ Statistical assumptions<br>✗ Simple designs |
| ✓ Intuitive approach<br>✓ Increased statistical flexibility<br>✓ Moderators | **IPD mega-trial** | ✗ **Biased results may be provided**<br>✗ Statistical assumptions<br>✗ Dependent on design, response variable |
| ✓ **Increased statistical flexibility**<br>✓ Moderators and heterogeneity | **IPD stratified** | ✗ Statistical assumptions<br>✗ Dependent on design, response variable<br>✗ Complexity |
| ✓ Independent of design, metric and statistical test | **Aggregation of $p$-values** | ✗ **Provides no effect size**<br>✗ $p$-value dependent on sample size and effect size |

According to Gomez et al. [2], different dimensions can be changed across experiments:

- **Operationalization**: refers to the operationalization of the treatments, metrics and measurement procedures used within the experiments (e.g., response variable scales and the use of test cases or experts to score participants' solutions).
- **Population**: refers to the characteristics of the participants within the experiments (e.g., students vs. professionals, different skills and different backgrounds).
- **Protocol**: refers to the "apparatus, materials, experimental objects, forms and procedures" used within the experiments (e.g., experimental tasks, experimental session length, and training duration).
- **Experimenters**: refers to the personnel involved in the experiments (e.g., the trainer, the measurer, and the analyst).

Table 7 shows a map between the families that we identified (first column), the aggregation techniques used (second column, where the aggregation techniques are [N]arrative synthesis, Aggregation of [P]-values, [A]D, IPD [M]ega-trial or IPD [S]tratified), the dimensions that changed across the experiments (third column, where the dimensions changed are [O]perationalization, [P]opulation, [Pr]otocol or [E]xperimenters), and other information that we will discuss later.

The population dimension is the one that varies most frequently within families. Populations are commonly changed across experiments to increase the external validity of the results [P21], [P13], [P1]. For example, Porter and Votta [P21] claimed that one of the goals of their family was to "...extend the external credibility of our results by studying professionals developers...". Unfortunately, introducing population changes across experiments may affect the population effect size being estimated and thus impact the validity of the conclusion [2]. For example, a treatment could be effective for students but not for professionals (or vice versa). In such circumstances, as each experiment is estimating a potentially different effect size and because sample sizes are usually small in SE, an "amalgamation" of potentially unreliable effect size estimates is offered as a joint result within families—if a fixed-effects model such as ANOVA, linear regression or fixed-effects meta-analysis models are used [21], [20], [49] (as is commonly the case within families). This issue becomes increasingly relevant as the number of changes introduced across the experiments increases [1] and as the sample size decreases [50].

We do not observe any relationship between the number of dimensions changed within families and the frequencies of use of each aggregation technique. For example, we expected to find that stricter techniques (such as IPD mega-trial or stratified) would be less used than others (e.g., Narrative synthesis) to analyze families with many dimension changes. Although the number of IPD mega-trial analyses conducted in such cases is small (only five families with three or four dimension changes used IPD mega-trial), almost all IPD stratified analyses were conducted in families with three to four dimension changes across the experiments. We consider this to be preliminary evidence suggesting that researchers do not to follow a procedure for selecting aggregation techniques based on the characteristics of the family. This result agrees with what was found in medicine years ago [28].

We identified several other elements that may also have influenced the selection of the aggregation technique/s. The rest of the columns in Table 7 present these elements. Let us examine them one by one.

Response variable changes appear in the fourth column of Table 7. As IPD cannot analyze experiments with different response variables, changing response variables' operationalizations may hinder the application of IPD—and at the same time increase the appeal of other techniques (e.g., AD with standardized effect sizes such as Cohen's d). According to Table 7, response variables' operationalizations rarely change across experiments within families (i.e., in just 20% of the families, including families where we lack information). As expected, whenever response variables' operationalizations changed, AD, Narrative synthesis or Aggregation of $p$-values were used to aggregate the experiments' results. Although Narrative synthesis and Aggregation of $p$-values look appealing in these circumstances, we recommend AD (as did Scanniello et al. [P1]), as AD *transparently* weights each experiment to produce the joint result and thus increases the reliability and transparency of the results. In addition, AD simultaneously provides an effect size and a $p$-value, and thus, allows to assess both the relevance and the significance of results [35]—contrary to Aggregation of $p$-values, which provides only a $p$-value, and Narrative synthesis, which does not provide a quantitative summary of the results.

Experimental design changes appear in the fifth column of Table 7. As some IPD statistical models cannot analyze groups of experiments with different experimental designs (e.g., repeated-measures ANOVA can analyze experiments with only within-subjects designs [27]), introducing changes across experiments' designs may hinder the application of IPD and favor the application of other more "flexible" techniques (e.g., Narrative synthesis, AD or Aggregation of $p$-values). As shown in Table 7, design changes were rarely introduced across experiments within families (in only 15% of the families). In 66% of the families introducing design changes, Narrative synthesis was used. For example, Juristo et al. [P2] analyzed each experiment independently with an ANOVA model with four factors (i.e., Technique, Program, Version and Fault) and then classified each factor into one of three categories (i.e., non-significant, significant or "doubtful"), depending on the number of times each factor was significant across the experiments. Despite its appeal, Narrative synthesis may be especially dangerous given the small sample sizes common in SE experiments [51] and thus the large variability of results expected due to natural variation [52]. For example, if two exact replications, each with a statistical power of 30%, are conducted (thus, there is a 30% probability of obtaining statistically significant results in each), there is a 0.09 (i.e., 0.3*0.3) probability of achieving two statistically significant results and a 0.49 (i.e., 0.7*0.7) probability of obtaining two non-significant results. In turn, there is a probability of 0.42 (i.e., 1-0.49-0.09) of obtaining one significant and one non-significant result and claiming conflicting results when in reality, both experiments' estimate exactly the same population effect size [36], [14]. In addition, finding two non-significant results across two experiments does not imply that the joint result is not statistically significant [21]. Simply that larger sample sizes are required to achieve statistical significance in each individual experiment [36].

The analysis technique/s used to analyze each experiment individually within families are shown in the last column of Table 7 (i.e., [A]NOVA, [N]on-parametric, [T]-test, [C]orrelation, [R]egression, [O]thers or None (-)). We assessed the techniques used to analyze individual experiments within families as we think that it may be tempting to use the same technique to analyze the family as a whole (e.g., by simply pooling the raw data of all experiments together and then analyzing them as coming from a single "big" experiment), and in turn, this may lead to a over-representation of IPD mega-trial.

As shown in Table 7, non-parametric statistical tests (e.g., Mann-Whitney U and Wilcoxon [27]) are frequently used to analyze individual experiments within families (in 51% of the families). The most common reason for using non-parametric tests appears to be the lack of normality of the data. For example, Fernandez et al. [P20] claimed that they used the Wilcoxon test to analyze each experiment individually as "...in most cases the data were not normal...", while Cruz-Lemus et al. [P28] claimed that they used the Kruskal-Wallis test as "... *Kruskal-Wallis is the most appropriate test... when there is non-normal distribution of the data...*". Among the families relying on non-parametric tests to analyze individual experiments, 62% rely on the same statistical test to analyze the family as a whole, regardless of the overall sample size achieved at the family level. For example, Hadar et al. [P33] used the Mann-Whitney test to analyze each individual experiment and the family as a whole, despite obtaining a sample size of approximately 80 subjects when the raw data were pooled together. Although this procedure appears to be consistent, it may not be optimal, as traditional statistical tests, such as ANOVA, are robust to departures from normality [53] (especially when the sample sizes are large, as when the raw data of all the experiments are pooled [54]), and they should be preferred over their non-parametric counterparts when the sample sizes get large [55]. We suggest that this large reliance on IPD mega-trial in SE may be due to the lack of normality of the data, the common use of non-parametric tests to analyze individual experiments, and the impossibility of accommodating factors other than "treatment" within the traditionally used non-parametric tests (e.g., Mann-Whitney U or Wilcoxon [27]).

## 6 RQ4: FAMILIES LIMITATIONS

We found that SE families share a series of common limitations with regard to joint data analysis.

For example, there is an over-reliance on Narrative synthesis to provide joint results. Narrative synthesis has been discouraged in mature experimental disciplines such as medicine and pharmacology due to its inability to provide a quantitative summary of the results and its subjectivity when providing joint results [11], [21]. Moreover, and as we discussed previously, Narrative synthesis fails to take into account the natural variation of results [21]. This issue may be especially relevant in SE, where small sample sizes are the norm rather than the exception [51], and thus, a large variability of results is expected [56]. As an example, and perhaps unknowingly, Narrative synthesis may have tricked Ceccato et al. [P10] into thinking that conflicting results materialized across two small experiments (with sample sizes of 13 each) when they claimed that "*...strangely enough, the trend observed... in Exp IV and V has alternating directions*" or Runeson et al. [P36] when they claimed that "*...the first replication was designed to be as exact as possible... Despite an attempt at an exact replication, the outcomes were not the same...*" in two experiments with sample sizes of 33 and 18. In particular, conflicting results (in terms of either $p$-values or effect sizes) may materialize simply because of the presence of small sample sizes, and thus, the large variation of results

TABLE 7
Changes introduced within families (ordered by total number of changes).

| ID | Techniques | Changes | Response | Design | Individual |
|---|---|---|---|---|---|
| [P2] | N,-,-,- | O,P,PR,E | ? | ✓ | A |
| [P4] | N,-,-,- | O,P,PR,E | ✗ | ✗ | C |
| [P5] | -,-,A,-,- | O,P,PR,E | ✗ | ✗ | C |
| [P6] | -,-,A,-,- | O,P,PR,E | ? | ✗ | A |
| [P31] | -,-,-,-,S | O,P,PR,E | ✗ | ✗ | N |
| [P36] | N,-,-,-,S | O,P,PR,E | ✗ | ✗ | N |
| [P1] | -,-,A,-,- | -,P,PR,E | ✓ | ✓ | - |
| [P3] | -,-,-,-,S | -,P,PR,E | ✗ | ✗ | R |
| [P7] | -,-,-,M,- | -,P,PR,E | ✗ | ✗ | - |
| [P8] | -,-,A,-,- | -,P,PR,E | ? | ✗ | N |
| [P9] | N,-,-,- | O,P,PR,- | ✓ | ✓ | A |
| [P11] | N,-,-,M,- | O,P,-,E | ✗ | ✗ | N,O |
| [P15] | -,-,-,M,- | -,P,PR,E | ✗ | ✗ | T |
| [P16] | N,-,-,-,S | -,P,PR,E | ✓ | ✗ | N |
| [P17] | -,-,A,-,- | -,P,PR,E | ✗ | ✗ | N,T |
| [P18] | -,-,A,-,- | -,P,PR,E | ✗ | ✗ | N |
| [P20] | N,-,A,-,- | -,P,PR,E | ? | ✗ | N |
| [P24] | -,P,A,-,- | -,P,PR,E | ✗ | ✗ | T |
| [P26] | -,-,-,M,S | -,P,PR,E | ✗ | ✗ | N |
| [P28] | -,-,A,M,- | -,P,PR,E | ✗ | ✗ | N |
| [P29] | N,P,-,-,- | O,P,PR,- | ✓ | ✓ | N |
| [P35] | N,-,A,-,- | O,P,PR,- | ✗ | ✗ | T |
| [P37] | N,-,-,- | -,P,PR,E | ✗ | ✓ | N |
| [P38] | -,-,A,-,- | O,P,PR,- | ✓ | ✗ | C |
| [P10] | N,-,-,M,- | -,P,-,E | ✗ | ✗ | N,O |
| [P12] | N,-,-,- | -,P,PR,- | ✗ | ✗ | N,T |
| [P13] | N,-,-,- | -,P,PR,- | ✗ | ✗ | C |
| [P14] | N,-,-,M, | -,P,-,E | ✗ | ✗ | N,T |
| [P19] | N,-,-,- | -,P,PR,- | ✗ | ✗ | N |
| [P22] | -,P,A,-,- | -,P,PR,- | ✗ | ✓ | T |
| [P25] | N,-,-,- | -,P,-,E | ✗ | ✗ | C |
| [P27] | -,-,-,M,- | -,P,-,E | ✗ | ✗ | - |
| [P30] | -,-,A,M,- | -,P,-,E | ✗ | ✗ | A |
| [P32] | N,-,-,- | -,P,PR,- | ✗ | ✗ | N |
| [P33] | -,-,A,M,- | -,P,PR,- | ✗ | ✗ | N |
| [P34] | -,-,-,M,- | -,P,-,E | ✗ | ✗ | N,T |
| [P21] | N,-,-,-,S | -,P,-,- | ✗ | ✗ | A |
| [P23] | -,-,-,M,- | -,-,PR,- | ✗ | ✗ | - |
| [P39] | -,-,A,M,- | -,P,-,- | ✗ | ✗ | N,A |

expected, and not because the experiments observe different realities [50], [14].

Narrative synthesis was used to assess moderator effects in 15% of the families. Again, despite its appeal, Narrative synthesis may be unreliable for detecting moderators, especially when the experiments are small. In particular, and as a large variability in the results is expected for small sample sizes [50], [14], there is a high risk of claiming that differences across experiments' results are due to moderator effects when in reality, such differences could have emerged just because of natural variation of results. For example, observing that one experiment provided different results than the rest, Ali et al. [P37] claimed that *"...One plausible explanation is that participants had more experience with standard UML state machines..."*. On its side, and after noticing that different results were obtained across two experiments within the family, Juristo et al. [P2] claimed that *"subjects might swap information about the programs and their faults at the end of each session. As a result of copying, the techniques applied... could be more effective at UdS and UPV than at other sites"*. Although such claims may add to the discussion, they may also be misleading. Unfortunately, Narrative synthesis does not allow to determine how much of the difference in the results is due to natural variation and how much is due to moderators [50], [14]. This may have affected the reliability of the conclusions reached within families applying Narrative synthesis to elicit moderators.

With regard to the use of AD, we have noticed some inconsistencies. For example, in some families, non-parametric statistical tests (e.g., Mann-Whitney U [27]) were used to analyze individual experiments—as according to the authors, the data did not follow normality. Then, parametric effect sizes (such as Cohen's d or Hedges' g) were calculated for being integrated with AD. As an example, Fernandez et al. [P39] analyzed each individual experiment by means of the Mann-Whitney U test (as the data did not follow normality) and then computed the Hedges' g of each experiment to provide a joint result by means of AD. Unfortunately, the use of parametric effect sizes (such as Cohen's d or Hedges' g) comes also at the cost of checking the statistical assumptions on which they are built (e.g., normality and homogeneity of variance [39], [38], [57]). This verification is more relevant for small sample sizes [38]—as those common in SE experiments. Thus, we suggest that if non-parametric tests are used to analyze individual experiments, at least for consistency, non-parametric effect sizes, such as Cliff's delta, should be used to provide joint results with AD [39], [38], [57].

Also with regard to the use of AD, we have noticed that despite the multiple changes usually introduced across experiments within families (and thus, the potential heterogeneity of results inadvertently introduced within families), fixed-effects models were commonly preferred over random-effects models—perhaps because of their 'on-average' smaller $p$-values [21], [20], and in turn, their "more significant" results. For example, Scanniello et al. [P1] analyzed a group of five replications with a fixed-effects meta-analysis model despite obtaining a relatively large and statistically significant heterogeneity of results, and Cruz et al. [P6] analyzed a group of nine effect sizes with fixed-effects models despite the observable heterogeneity in the forest plot. Unfortunately, not acknowledging the heterogeneity of results during the statistical analysis may limit the reliability of the joint results [21], which may be worrisome, especially if a large number of changes have been introduced across experiments.

With regard to the use of IPD mega-trial to analyze families, and as previously discussed in medicine [43], [42], IPD mega-trial may provide biased results if subjects within experiments more closely resemble each other that do subjects across experiments (e.g., when some experiments are run with professionals and others with students), or if data are unbalanced across treatments and experiments (e.g., when some experiments are larger than others or when missing data materializes in some experiments but not in others). Given that missing data is common in SE due to protocol deviators or drop-outs [58], that families of experiments are usually comprised by experiments with professionals and students, and that experiments run with professionals tend to be smaller than those run with students, we are skeptical about the suitability of IPD mega-trial to analyze SE families.

IPD mega-trial was used to elicit moderators in 18% of the families [P10], [P11], [P28], [P26], [P33], [P30], [P15]. In such families, experiments are commonly performed with different types of subjects, and a certain "tag" is used to represent each type of subject as if it was the "Experiment" factor in an IPD stratified analysis. For example, experience (i.e., PhD, master, undergraduate) was used by Ricca et al. and Ceccato et al. [P26], [P14] to represent the "Experiment" factor in an IPD stratified model. Despite its intuitiveness, this approach may be misleading: differences across experiments' results may not be due to single moderators, especially because other unknown variables (e.g., age, motivation, skills, treatment conformance, the materialization of threats to validity in some experiments and not in others, etc. [59]), or confounding factors (e.g., if more than one change were purposefully introduced simultaneously across the experiments) may also be causes of the differences in the results [60], [61].

Finally, we would also like to comment on the ability of families to increase statistical power [3] and the extent to which this is dependent upon the "severity" of the changes introduced across the experiments within families. Specifically, changes to a certain dimension—or multiple dimensions—across experiments may introduce heterogeneity of results [62]. Under such circumstances, random-effects models should be preferred over fixed-effects models [21], [20]. This may have a noticeable impact on statistical power (as random-effects models tend to be more conservative than their fixed-effects counterparts [12], [21], [63]). Thus, if changes are introduced across experiments, larger sample sizes—and potentially a larger number of experiments—may be needed to reach the same significance level of fixed-effects models [21]. Therefore, introducing many changes across experiments may be detrimental to families' statistical power.

## 7 GOOD PRACTICES FOR ANALYZING FAMILIES

We have developed a preliminary list of recommendations to analyze families of experiments based on the common limitations we observed with regard to joint data analysis. These recommendations follow:

- **Avoid Narrative synthesis if possible.** Despite its appeal, the application of Narrative synthesis is dangerous given current sample size limitations—and thus, the large variability of results expected—in SE experiments. In addition, Narrative synthesis provides only a textual summary of the results rather than a quantitative summary of results (such as a $p$-value or effect size). Thus, the application of Narrative synthesis to aggregate experiments' results goes against the best practices of SE experimentation [17], [64], [65], [66], [7], where the reporting of $p$-values and effect sizes to summarize results is encouraged.
- **Avoid IPD mega-trial if possible.** Although intuitive, IPD mega-trial fails to account for the heterogeneity of results across experiments. Unfortunately, heterogeneity may materialize if changes are introduced—deliberately or inadvertently—across experiments. IPD mega-trial also fails to account for the plausible correlation of participants' scores within experiments (as subjects within an experiment may more closely resemble each other than do subjects across experiments) and the existence of experiments with different sample sizes and missing data [43], [42].
- **Avoid Aggregation of $p$-values if possible.** In Aggregation of $p$-values, each experiment contributes identically to the overall conclusion, independent of its quality, effect size or sample size. Moreover, Aggregation of $p$-values does not provide an effect size, which hinders the assessment of the relevance of the results (e.g., how *large* is the joint effect? [21], [20]). In view that experiments within families usually have different sample sizes, and that providing an assessment of the relevance of results is key to argue about the performance of SE technologies [35], we are skeptical about the applicability of Aggregation of $p$-values to analyze SE families.
- **When data do not follow normality**. The robustness of traditional parametric statistical tests, such as ANOVAs or $t$-tests, to departures from normality has been assessed repeatedly, even for sample sizes smaller than those typical in SE experiments [67], [53], [55], [68], [69]. The robustness of traditionally used parametric statistical tests to departures from normality is even greater for sample sizes in the hundreds [54]—such as those resulting from pooling the raw data of all the experiments within a family. The superiority of parametric statistical tests over non-parametric tests, such as the Mann-Whitney U or Wilcoxon, is widely acknowledged [70], [67], [53], especially with regard to the *interpretability of the results* (as the results can be presented in natural units, e.g., differences between means, instead of differences between mean ranks) and *statistical flexibility* (e.g., multiple factors such as type of subject and experiment can be included). Although we concede that defenders of both approaches (i.e., parametric vs. non-parametric) can be found in the literature [71], [72], [70], [67], [53], [55], [68], [69], we believe that a good compromise is that followed by the authors of some families of experiments (e.g., Laitenberger et al. [P22] or Pfahl et al. [P24]). Specifically, from the perspective that the Wilcoxon signed-rank test and the dependent $t$-test provided similar results, Laitenberger et al. [P22] used the results of the $t$-test. If, despite the robustness of parametric statistical tests to departures from normality, other approaches want to be followed, we recommend some of the following:

- **AD.** Use either (1) *non-parametric effect sizes* (e.g., Cliff's delta or probability of superiority [39]) or (2) *bootstrap* to obtain the standard error of the selected parametric effect size (e.g., Cohen's d or Hedges' g) [73].
- **IPD stratified.** Use either (1) *a permutation test* (following an approach similar to that of Ricca et al. [P16]); (2) a *Generalized Linear Model* to accommodate the distribution of the data (e.g., by means of logistic regression [74], [75]); or (3) use *bootstrap* [76], [77].

• **Check the statistical assumptions of AD and IPD.** If sample sizes are small and the data are not normal or have different variances, parametric effect sizes (such as Cohen's d or Hedges' g [21]) may be unreliable [39], [38], [57]. Similarly, if an IPD stratified statistical model (e.g., ANOVA) is used, check its statistical assumptions (e.g., normality or homogeneity of variances) or consult specialized literature to investigate about the robustness of statistical tests to departure from their assumptions [72], [70], [53], [55], [68], [69].

• **Acknowledge the heterogeneity of the results** when providing joint results. If changes are introduced across experiments (e.g., different populations or experimental designs), such changes may impact the effect size being estimated in each experiment [21]. As other researchers have previously, we recommend the application of random-effects models rather than fixed-effects models, as the former reduce to the latter if no heterogeneity of results materializes [21].

• **Plan ahead the analysis procedure to follow.** We encourage researchers conducting families of experiments to plan ahead *at least* the pre-processing steps that will be performed (e.g., the procedure used to remove outliers and graph data), the statistical tests that will be used (e.g., are researchers interested in differences between means (and thus parametric tests)), and the aggregation techniques to be used (e.g., are researchers interested in providing intuitive visualizations of the results, such as with AD, or are researchers interested in evaluating results in natural units, such as with IPD). If the application of such procedures, tests or aggregation techniques is infeasible after conducting the experiments (e.g., due to unexpected data losses or experimental restrictions that forced certain design changes), the rationale of such changes and their impact on the planned analysis procedures should be discussed before presenting the joint results or eliciting moderators.

## 8 GOOD PRACTICES FOR REPORTING FAMILIES' RESULTS

While conducting this SMS, we found that assessing the appropriateness of the aggregation technique/s used within families was not straightforward. This difficulty was in part because the articles left out some relevant information (e.g., are the raw data of all the experiments available?, are the response variables' operationalizations identical across the experiments?) and in part because the raw data of the families of experiments were not made public—and thus, reanalyzing the family with each technique to check their suitability was infeasible. Therefore, we have extracted a series of elements that we believe are relevant for assessing the reliability of joint results and the suitability of the aggregation technique/s applied. We suggest reporting at least the following elements in research articles on families of experiments:

• **Raw data availability.** Report whether the raw data are available at the time the data are analyzed. If the raw data are available, techniques such as IPD stratified or AD may be the preferred option to provide joint results or elicit moderators.

• **Response variable changes.** Report whether the same response variable operationalization is used across the experiments, as if this is the case, there is no need to compute standardized effect sizes (e.g., Cohen's d) to conduct AD. In this situation, either IPD stratified or AD with *unstandardized* units can be used to increase the interpretability of the results [21], [40], [44].

• **Experiments' sample sizes.** Report the number of subjects that participated in each experiment since this information is useful for computing various effect sizes [21], and arguing about the generalizability of the results.

• **The relationship between the participants across the experiments.** Report if different subjects participated in each experiment. If the participants are the same, and this was not accounted for by the aggregation technique used, the joint results may be misleading.

• **The technique/s used to analyze the family of experiments.** Motivate the selection of the aggregation technique/s used and why such technique/s were selected given the characteristics of the family of experiments.

- **The elements that changed across the experiments.** Explicitly state the changes introduced across experiments so that it is possible to assess the suitability of the aggregation technique/s applied.

Finally, we observed that the raw data of the families were not accessible in most cases—despite 6 of 39 families claiming to have published the raw data [P14], [P35][P8], [P17], [P19], [P1]. The unavailability of the raw data precludes re-analysis with other—perhaps more suitable—techniques to verify the robustness of the results, and hinders the reproducibility of the original results. Thus, if possible, **we encourage researchers reporting families of experiments to make their raw data available**.

## 9 THREATS TO VALIDITY

In this section, we discuss the main threats to the validity of our SMS according to Petersen et al.'s guidelines [15].

When conducting a secondary study, the search terms should identify as many relevant papers as possible to provide an accurate overview of the topic and its structure [15]. We piloted different terms and search strings to reduce the risk of missing relevant primary studies. Due to the inconsistency in the terms that we used to refer to families of experiments in SE, we used synonyms of the terms obtained from a reference set of five articles [P38], [P14], [P8], [P2], [P36]. This process helped us to broaden the results of our SMS and increase the reliability of the findings.

We had to make a trade-off between the complexity of the search string (to not be too restrictive) and the looseness of the terms (to not be too liberal). In particular, the words "experiment" and "aggregation", "group of studies" and "series of experiments" appeared in many different disciplines unrelated to the scope of our research, which added a large amount of noise to our results. Therefore, we had to restrict the search space to relevant venues on SE experimentation. We do not believe this restriction had a major impact on the results, as the techniques that we found in our primary studies were similar to those used in mature disciplines such as medicine and pharmacology [20], [21], [62].

To increase the precision of the results, we confined the search space to the venues surveyed in a well-known secondary study on SE experiments [78]. We acknowledge the possibility of publication bias in the results (as we restricted our search to well-known venues for publishing empirical research). Although publication bias may have conditioned our results, we complemented the search with a backward snowballing process [17].

After identifying all the articles, the selection of primary studies within the scope of research is crucial for reaching relevant conclusions. We defined an explicit exclusion criteria to minimize subjectivity and to prevent the omission of valid articles. This exclusion criteria was applied by the first two researchers to select valid primary studies. In the event of disagreement, a third researcher helped with the inclusion or exclusion decision.

We acknowledge that other researchers may have gathered a different list of references for portraying the advantages and disadvantages of the aggregation techniques that we identified. Unfortunately, we could not collect such references systematically due to the enormous list of references prompted by online databases with the terms "aggregation", "experiments", "data" or "effect sizes". However, we tried to minimize this shortcoming by considering the comments made on the aggregation techniques on well-known references in mature experimental disciplines [62], [79], seminal works on vote-counting and narrative synthesis [80], [81], and meta-analysis and reproducibility of results [21], [56], [14].

In addition, and although we used the same naming conventions as those used in medicine to categorize the aggregation techniques (e.g., AD, IPD), this does not imply that those techniques have been applied just in medicine: the same aggregation techniques—albeit with different names—have also been applied in disciplines such as econometrics, ecology, biology, and social sciences, just to name a few [21], [20], [82]. We relied on this discipline and not others, as SE experimental research has previously imported from there the data analysis techniques used to analyze individual experiments [83], [64], well-known research methods to conduct literature reviews (e.g., the procedures for conducting SLRs [7]) or empirical research concepts (e.g., Evidence Based Software Engineering [66]) among others.

Finally, the raw data were accessible in only 7% of the families of experiments that we identified. This unavailability of raw data hindered re-analysis and, thus, a formal assessment of the suitability of the techniques applied to analyze both individual experiments and the families as a whole. Therefore, providing finely tailored recommendations about the suitability of the statistical techniques applied to analyze each family was infeasible.

## 10 CONCLUSION

Families of experiments (i.e., groups of interrelated experiments with the same goal [1]) are being run in increasing numbers in SE. The characteristics, implications and particularities of families compared to other study types (such as groups of experiments gathered by means of SLRs) are yet to be defined in SE. Fine-tuning the term family of experiments may help SE researchers to apply consistent techniques to analyze families.

We define a family of experiments as a group of replications where access to the raw data is guaranteed, the settings of the experiments are known by the researchers, and at least three experiments evaluate the effects of at least two different technologies on the same response variable. Several techniques have been used to aggregate experiments' results within families in SE: Narrative synthesis, AD, IPD mega-trial, IPD stratified and Aggregation of $p$-values. The rationale used to select the aggregation technique/s within families is rarely discussed in research articles. The technique used to analyze families may impact the reliability of the joint results and undermine families' potential to elicit moderators.

A total of 46% of the families used Narrative synthesis to obtain joint conclusions or study moderators. Narrative synthesis may be especially misleading for aggregating experiments' results within families due to their small sample sizes. A total of 38% of the families used AD. Some of the families applying AD resorted to parametric effect sizes (e.g., Cohen's d or Hedges' g) despite acknowledging small sample sizes and data that did not follow normality. This may have impacted the reliability of the findings in such families. 33% of families used IPD mega-trial. Unbalanced data and the multiple changes commonly introduced across experiments within families may have led to unreliable results with IPD mega-trial. A total of 15% of the families used IPD stratified to aggregate the experiments' results following procedures similar to those followed in medicine [20]. Finally, Aggregation of $p$-values was used in only 7% of the families. Aggregation of $p$-values identically weights the contribution of each experiment to the joint result and does not provide a joint effect size. This may have limited the interpretability of the results in families applying Aggregation of $p$-values.

The aggregation technique/s used within families may impact the reliability of the joint results. Reliable aggregation techniques, such as AD or IPD stratified, appear to be suitable to analyze SE families. However, their use comes at a cost: statistical assumptions need to be checked before interpreting joint results. We recommend minimizing the use of Narrative synthesis as it does not provide a quantitative summary of results, involves subjective judgment, and does not take advantage of the raw data when producing joint results or eliciting moderators. We recommend minimizing the use of Aggregation of $p$-values as it does not provide any method for assessing the relevance of the results, and in its basic form, identically weights the contribution of each experiment to the joint conclusion, regardless of effect size, sample size, quality and experimental design. IPD mega-trial should also be avoided as it may provide biased results if the data are unbalanced across or within experiments or if subjects with an experiment more closely resemble each other than do subjects across experiments.

Finally, we are observing a valuable increase in the number of experiments being included within SE families (currently an average of five per family). This may lead the SE community to a next stage of maturity. However, reporting deficiencies and the blind application of unsuitable aggregation techniques could limit the reliability of joint results and moderator effects within families. Researchers should assess the suitability of the aggregation techniques applied and provide a rationale for why such techniques were selected to analyze their respective families. Finally, we urge researchers conducting families of experiments in SE to plan ahead their data analyses and improve the transparency of their reports with respect to data, statistical assumptions and data analysis techniques.

## ACKNOWLEDGMENTS

This research was developed with the support of the Spanish Ministry of Science and Innovation project TIN2014-60490-P. We would like to thank the anonymous reviewers who helped to polish this research article.

## REFERENCES

[1] V. R. Basili, F. Shull, and F. Lanubile, "Building knowledge through families of experiments," *Software Engineering, IEEE Transactions on*, vol. 25, no. 4, pp. 456–473, 1999.

[2] O. S. Gomez, N. Juristo, and S. Vegas, "Understanding replication of experiments in software engineering: A classification," *Information and Software Technology*, vol. 56, no. 8, pp. 1033–1048, 2014.

[3] E. Fernández, O. Dieste, P. M. Pesado, and R. García Martínez, "The importance of using empirical evidence in software engineering," 2011.

[4] J. C. Carver, N. Juristo, M. T. Baldassarre, and S. Vegas, "Introduction to special issue on replications of software engineering experiments," *Empirical Softw. Engg.*, vol. 19, no. 2, pp. 267–276, Apr. 2014.

[5] M. Ciolkowski, F. Shull, and S. Biffl, "A family of experiments to investigate the influence of context on the effect of inspection techniques," *Proceedings of the Empirical Assessment in Software Engineering, IEEE*, 2002.

[6] N. Juristo and S. Vegas, "Using differences among replications of software engineering experiments to gain knowledge," in *Proceedings of the 2009 3rd International Symposium on Empirical Software Engineering and Measurement*. IEEE Computer Society, 2009, pp. 356–366.

[7] B. Kitchenham, "Procedures for performing systematic reviews," *Keele, UK, Keele University*, vol. 33, no. 2004, pp. 1–26, 2004.

[8] N. Juristo and S. Vegas, "The role of non-exact replications in software engineering experiments," *Empirical Software Engineering*, vol. 16, no. 3, pp. 295–324, 2011.

[9] F. J. Shull, J. C. Carver, S. Vegas, and N. Juristo, "The role of replications in empirical software engineering," *Empirical Software Engineering*, vol. 13, no. 2, pp. 211–218, 2008.

[10] B. Kitchenham, "The role of replications in empirical software engineeringa word of warning," *Empirical Software Engineering*, vol. 13, no. 2, pp. 219–221, 2008.

[11] H. Cooper and E. A. Patall, "The relative benefits of meta-analysis conducted with individual participant data versus aggregated data." *Psychological methods*, vol. 14, no. 2, p. 165, 2009.

[12] J. Lau, J. P. Ioannidis, and C. H. Schmid, "Summing up evidence: one answer is not always enough," *The lancet*, vol. 351, no. 9096, pp. 123–127, 1998.

[13] A. Haidich, "Meta-analysis in medical research," *Hippokratia*, vol. 14, no. 1, pp. 29–37, 2011.

[14] J. Ioannidis, N. Patsopoulos, and H. Rothstein, "Research methodology: reasons or excuses for avoiding meta-analysis in forest plots," *BMJ: British Medical Journal*, vol. 336, no. 7658, pp. 1413–1415, 2008.

[15] K. Petersen, R. Feldt, S. Mujtaba, and M. Mattsson, "Systematic mapping studies in software engineering," in *12th international conference on evaluation and assessment in software engineering*, vol. 17, no. 1. sn, 2008, pp. 1–10.

[16] B. Kitchenham and S. Charters, "Guidelines for performing systematic literature reviews in software engineering," 2007.

[17] C. Wohlin, "Guidelines for snowballing in systematic literature studies and a replication in software engineering," in *Proceedings of the 18th International Conference on Evaluation and Assessment in Software Engineering*. ACM, 2014, p. 38.

[18] F. Shull, M. G. Mendonçça, V. Basili, J. Carver, J. C. Maldonado, S. Fabbri, G. H. Travassos, and M. C. Ferreira, "Knowledge-sharing issues in experimental software engineering," *Empirical Software Engineering*, vol. 9, no. 1-2, pp. 111–137, 2004.

[19] J. C. Carver, N. Juristo, M. T. Baldassarre, and S. Vegas, "Replications of software engineering experiments," 2014.

[20] A. Whitehead, *Meta-analysis of controlled clinical trials*. John Wiley & Sons, 2002, vol. 7.

[21] M. Borenstein, L. V. Hedges, J. P. Higgins, and H. R. Rothstein, *Introduction to Meta-Analysis*. John Wiley & Sons, 2011.

[22] D. Jackson, R. Riley, and I. R. White, "Multivariate meta-analysis: Potential and promise," *Statistics in medicine*, vol. 30, no. 20, pp. 2481–2498, 2011.

[23] C. Anello, R. T. ONeill, and S. Dubey, "Multicentre trials: a us regulatory perspective," *Statistical Methods in Medical Research*, vol. 14, no. 3, pp. 303–318, 2005.

[24] J. A. Lewis, "Statistical principles for clinical trials (ich e9): an introductory note on an international guideline," *Statistics in medicine*, vol. 18, no. 15, pp. 1903–1942, 1999.

[25] L. A. Stewart, M. Clarke, M. Rovers, R. D. Riley, M. Simmonds, G. Stewart, and J. F. Tierney, "Preferred reporting items for a systematic review and meta-analysis of individual participant data: the prisma-ipd statement," *Jama*, vol. 313, no. 16, pp. 1657–1665, 2015.

[26] L. Bero and D. Rennie, "The cochrane collaboration: preparing, maintaining, and disseminating systematic reviews of the effects of health care," *Jama*, vol. 274, no. 24, pp. 1935–1938, 1995.

[27] A. Field, *Discovering statistics using IBM SPSS statistics*. Sage, 2013.

[28] M. C. Simmonds, J. P. Higginsa, L. A. Stewartb, J. F. Tierneyb, M. J. Clarke, and S. G. Thompson, "Meta-analysis of individual patient data from randomized trials: a review of methods used in practice," *Clinical Trials*, vol. 2, no. 3, pp. 209–217, 2005.

[29] J. Popay, H. Roberts, A. Sowden, M. Petticrew, L. Arai, M. Rodgers, N. Britten, K. Roen, and S. Duffy, "Guidance on the conduct of narrative synthesis in systematic reviews," *A product from the ESRC methods programme Version*, vol. 1, p. b92, 2006.

[30] M. Rodgers, A. Sowden, M. Petticrew, L. Arai, H. Roberts, N. Britten, and J. Popay, "Testing methodological guidance on the conduct of narrative synthesis in systematic reviews: effectiveness of interventions to promote smoke alarm ownership and function," *Evaluation*, vol. 15, no. 1, pp. 49–73, 2009.

[31] R. Rosenthal, H. Cooper, and L. Hedges, "Parametric measures of effect size," *The handbook of research synthesis*, pp. 231–244, 1994.

[32] S. Nakagawa, "A farewell to bonferroni: the problems of low statistical power and publication bias," *Behavioral Ecology*, vol. 15, no. 6, pp. 1044–1045, 2004.

[33] R. Coe, "It's the effect size, stupid: What effect size is and why it is important," 2002.

[34] J. Cohen, "Statistical power analysis for the behavioral sciences lawrence earlbaum associates," *Hillsdale, NJ*, pp. 20–26, 1988.

[35] V. B. Kampenes, T. Dybå, J. E. Hannay, and D. I. Sjøberg, "A systematic review of effect size in software engineering experiments," *Information and Software Technology*, vol. 49, no. 11, pp. 1073–1086, 2007.

[36] G. Cumming, *Understanding the new statistics: Effect sizes, confidence intervals, and meta-analysis*. Routledge, 2013.

[37] M. R. Hess and J. D. Kromrey, "Robust confidence intervals for effect sizes: A comparative study of cohensd and cliffs delta under non-normality and heterogeneous variances," in *Annual Meeting of the American Educational Research Association*, 2004, pp. 12–16.

[38] C.-Y. J. Peng and L.-T. Chen, "Beyond cohen's d: Alternative effect size measures for between-subject designs," *The Journal of Experimental Education*, vol. 82, no. 1, pp. 22–50, 2014.

[39] C. O. Fritz, P. E. Morris, and J. J. Richler, "Effect size estimates: current use, calculations, and interpretation." *Journal of Experimental Psychology: General*, vol. 141, no. 1, p. 2, 2012.

[40] L. A. Stewart and J. F. Tierney, "To ipd or not to ipd? advantages and disadvantages of systematic reviews using individual patient data," *Evaluation & the health professions*, vol. 25, no. 1, pp. 76–97, 2002.

[41] H. Quené and H. Van den Bergh, "On multi-level modeling of data from repeated measures designs: A tutorial," *Speech Communication*, vol. 43, no. 1-2, pp. 103–121, 2004.

[42] G. Abo-Zaid, B. Guo, J. J. Deeks, T. P. Debray, E. W. Steyerberg, K. G. Moons, and R. D. Riley, "Individual participant data meta-analyses should not ignore clustering," *Journal of clinical epidemiology*, vol. 66, no. 8, pp. 865–873, 2013.

[43] H. C. Kraemer, "Pitfalls of multisite randomized clinical trials of efficacy and effectiveness," *Schizophrenia Bulletin*, vol. 26, no. 3, pp. 533–541, 2000.

[44] T. Debray, K. G. Moons, G. Valkenhoef, O. Efthimiou, N. Hummel, R. H. Groenwold, and J. B. Reitsma, "Get real in individual participant data (ipd) meta-analysis: a review of the methodology," *Research synthesis methods*, vol. 6, no. 4, pp. 293–309, 2015.

[45] G. B. Stewart, D. G. Altman, L. M. Askie, L. Duley, M. C. Simmonds, and L. A. Stewart, "Statistical analysis of individual participant data meta-analyses: a comparison of methods and recommendations for practice," *PloS one*, vol. 7, no. 10, p. e46042, 2012.

[46] A. Birnbaum, "Combining independent tests of significance," *Journal of the American Statistical Association*, vol. 49, no. 267, pp. 559–574, 1954.

[47] G. Leandro, *Meta-analysis in Medical Research: The handbook for the understanding and practice of meta-analysis*. John Wiley & Sons, 2008.

[48] J. Miller, "Applying meta-analytical procedures to software engineering experiments," *Journal of Systems and Software*, vol. 54, no. 1, pp. 29–39, 2000.

[49] H. Brown and R. Prescott, *Applied mixed models in medicine*. John Wiley & Sons, 2014.

[50] K. S. Button, J. P. Ioannidis, C. Mokrysz, B. A. Nosek, J. Flint, E. S. Robinson, and M. R. Munafò, "Power failure: why small sample size undermines the reliability of neuroscience," *Nature Reviews Neuroscience*, vol. 14, no. 5, p. 365, 2013.

[51] T. Dybå, V. B. Kampenes, and D. I. Sjøberg, "A systematic review of statistical power in software engineering experiments," *Information and Software Technology*, vol. 48, no. 8, pp. 745–755, 2006.

[52] S. E. Maxwell, M. Y. Lau, and G. S. Howard, "Is psychology suffering from a replication crisis? what does failure to replicate really mean?" *American Psychologist*, vol. 70, no. 6, p. 487, 2015.

[53] A. J. Vickers, "Parametric versus non-parametric statistics in the analysis of randomized trials with non-normally distributed data," *BMC medical research methodology*, vol. 5, no. 1, p. 35, 2005.

[54] T. Lumley, P. Diehr, S. Emerson, and L. Chen, "The importance of the normality assumption in large public health data sets," *Annual review of public health*, vol. 23, no. 1, pp. 151–169, 2002.

[55] M. W. Fagerland, "t-tests, non-parametric tests, and large studiesa paradox of statistical practice?" *BMC Medical Research Methodology*, vol. 12, no. 1, p. 78, 2012.

[56] M. Egger, G. Davey-Smith, and D. Altman, *Systematic reviews in health care: meta-analysis in context*. John Wiley & Sons, 2008.

[57] G. Macbeth, E. Razumiejczyk, and R. D. Ledesma, "Cliff's delta calculator: A non-parametric effect size program for two groups of observations," *Universitas Psychologica*, vol. 10, no. 2, pp. 545–555, 2011.

[58] O. Dieste, G. Raura, P. Rodríguez *et al.*, "Professionals are not superman: failures beyond motivation in software experiments," in *Conducting Empirical Studies in Industry (CESI), 2017 IEEE/ACM 5th International Workshop on*. IEEE, 2017, pp. 27–32.

[59] O. Dieste, A. M. Aranda, F. Uyaguari, B. Turhan, A. Tosun, D. Fucci, M. Oivo, and N. Juristo, "Empirical evaluation of the effects of experience on code quality and programmer productivity: an exploratory study," *Empirical Software Engineering*, vol. 22, no. 5, pp. 2457–2542, 2017.

[60] R. H. Groenwold, A. R. T. Donders, G. J. van der Heijden, A. W. Hoes, and M. M. Rovers, "Confounding of subgroup analyses in randomized data," *Archives of internal medicine*, vol. 169, no. 16, pp. 1532–1534, 2009.

[61] B. Dijkman, B. Kooistra, and M. Bhandari, "How to work with a subgroup analysis," *Canadian Journal of Surgery*, vol. 52, no. 6, p. 515, 2009.

[62] J. P. Higgins and S. Green, *Cochrane handbook for systematic reviews of interventions*. John Wiley & Sons, 2011, vol. 4.

[63] V. Huta, "When to use hierarchical linear modeling," *Quant Methods Psychol*, vol. 10, no. 1, pp. 13–28, 2014.

[64] N. Juristo and A. M. Moreno, *Basics of software engineering experimentation*. Springer Science & Business Media, 2011.

[65] T. Dyba, B. A. Kitchenham, and M. Jorgensen, "Evidence-based software engineering for practitioners," *IEEE software*, vol. 22, no. 1, pp. 58–65, 2005.

[66] B. A. Kitchenham, T. Dyba, and M. Jorgensen, "Evidence-based software engineering," in *Proceedings of the 26th international conference on software engineering*. IEEE Computer Society, 2004, pp. 273–281.

[67] J. C. De Winter, "Using the student's t-test with extremely small sample sizes." *Practical Assessment, Research & Evaluation*, vol. 18, no. 10, 2013.

[68] E. Schmider, M. Ziegler, E. Danay, L. Beyer, and M. Bühner, "Is it really robust?" *Methodology*, 2010.

[69] U. Wadgave and M. R. Khairnar, "Parametric tests for likert scale: For and against," *Asian journal of psychiatry*, vol. 24, pp. 67–68, 2016.

[70] G. Norman, "Likert scales, levels of measurement and the laws of statistics," *Advances in health sciences education*, vol. 15, no. 5, pp. 625–632, 2010.

[71] A. Arcuri and L. Briand, "A practical guide for using statistical tests to assess randomized algorithms in software engineering," in *Software Engineering (ICSE), 2011 33rd International Conference on*. IEEE, 2011, pp. 1–10.

[72] E. Whitley and J. Ball, "Statistics review 6: Nonparametric methods," *Critical care*, vol. 6, no. 6, p. 509, 2002.

[73] S. Nakagawa and I. C. Cuthill, "Effect size, confidence interval and statistical significance: a practical guide for biologists," *Biological reviews*, vol. 82, no. 4, pp. 591–605, 2007.

[74] C. E. McCulloch and J. M. Neuhaus, *Generalized linear mixed models*. Wiley Online Library, 2001.

[75] B. M. Bolker, M. E. Brooks, C. J. Clark, S. W. Geange, J. R. Poulsen, M. H. H. Stevens, and J.-S. S. White, "Generalized linear mixed models: a practical guide for ecology and evolution," *Trends in ecology & evolution*, vol. 24, no. 3, pp. 127–135, 2009.

[76] S. Ren, H. Lai, W. Tong, M. Aminzadeh, X. Hou, and S. Lai, "Nonparametric bootstrapping for hierarchical data," *Journal of Applied Statistics*, vol. 37, no. 9, pp. 1487–1498, 2010.

[77] T. C. Hesterberg, "What teachers should know about the bootstrap: Resampling in the undergraduate statistics curriculum," *The American Statistician*, vol. 69, no. 4, pp. 371–386, 2015.

[78] D. I. Sjøberg, J. E. Hannay, O. Hansen, V. B. Kampenes, A. Karahasanovic, N.-K. Liborg, and A. C. Rekdal, "A survey of controlled experiments in software engineering," *Software Engineering, IEEE Transactions on*, vol. 31, no. 9, pp. 733–753, 2005.

[79] D. B. Petitti *et al.*, *Meta-analysis, decision analysis, and cost-effectiveness analysis: methods for quantitative synthesis in medicine.* OUP USA, 2000, no. 31.

[80] L. V. Hedges and I. Olkin, "Three vote-counting methods for the estimation of effect size and statistical significance of combined results," in *annual meeting of the American Research Association, San Francisco, California*, 1979.

[81] ——, "Vote-counting methods in research synthesis." *Psychological bulletin*, vol. 88, no. 2, p. 359, 1980.

[82] F. Harrison, "Getting started with meta-analysis," *Methods in Ecology and Evolution*, vol. 2, no. 1, pp. 1–10, 2011.

[83] C. Wohlin, P. Runeson, M. Höst, M. C. Ohlsson, B. Regnell, and A. Wesslén, *Experimentation in software engineering*. Springer Science & Business Media, 2012.

**PRIMARY STUDIES**

[P1] G. Scanniello, C. Gravino, G. Tortora, M. Genero, M. Risi, J. A. Cruz-Lemus, and G. Dodero, "Studying the effect of uml-based models on source-code comprehensibility: Results from a long-term investigation," in *International Conference on Product-Focused Software Process Improvement*. Springer, 2015, pp. 311–327.

[P2] N. Juristo, S. Vegas, M. Solari, S. Abrahao, and I. Ramos, "Comparing the effectiveness of equivalence partitioning, branch testing and code reading by stepwise abstraction applied by subjects," in *2012 IEEE Fifth International Conference on Software Testing, Verification and Validation*. IEEE, 2012, pp. 330–339.

[P3] J. L. Krein, L. Prechelt, N. Juristo, A. Nanthaamornphong, J. C. Carver, S. Vegas, C. D. Knutson, K. D. Seppi, and D. L. Eggett, "A multi-site joint replication of a design patterns experiment using moderator variables to generalize across contexts," *IEEE Transactions on Software Engineering*, vol. 42, no. 4, pp. 302–321, 2016.

[P4] G. Canfora, F. García, M. Piattini, F. Ruiz, and C. A. Visaggio, "A family of experiments to validate metrics for software process models," *Journal of Systems and Software*, vol. 77, no. 2, pp. 113–129, 2005.

[P5] M. E. Manso, J. A. Cruz-Lemus, M. Genero, and M. Piattini, "Empirical validation of measures for uml class diagrams: A meta-analysis study," in *International Conference on Model Driven Engineering Languages and Systems*. Springer, 2008, pp. 303–313.

[P6] J. A. Cruz-Lemus, M. Genero, M. E. Manso, S. Morasca, and M. Piattini, "Assessing the understandability of uml statechart diagrams with composite statesa family of empirical studies," *Empirical Software Engineering*, vol. 14, no. 6, pp. 685–719, 2009.

[P7] E. Figueiredo, A. Garcia, M. Maia, G. Ferreira, C. Nunes, and J. Whittle, "On the impact of crosscutting concern projection on code measurement," in *Proceedings of the tenth international conference on Aspect-oriented software development*. ACM, 2011, pp. 81–92.

[P8] S. Abrahao, C. Gravino, E. Insfran, G. Scanniello, and G. Tortora, "Assessing the effectiveness of sequence diagrams in the comprehension of functional requirements: Results from a family of five experiments," *IEEE Transactions on Software Engineering*, vol. 39, no. 3, pp. 327–342, 2013.

[P9] N. Salleh, E. Mendes, and J. Grundy, "Investigating the effects of personality traits on pair programming in a higher education setting through a family of experiments," *Empirical Software Engineering*, vol. 19, no. 3, pp. 714–752, 2014.

[P10] M. Ceccato, M. Di Penta, P. Falcarin, F. Ricca, M. Torchiano, and P. Tonella, "A family of experiments to assess the effectiveness and efficiency of source code obfuscation techniques," *Empirical Software Engineering*, vol. 19, no. 4, pp. 1040–1074, 2014.

[P11] M. Ceccato, A. Marchetto, L. Mariani, C. D. Nguyen, and P. Tonella, "Do automatically generated test cases make debugging easier? an experimental assessment of debugging effectiveness and efficiency," *ACM Transactions on Software Engineering and Methodology (TOSEM)*, vol. 25, no. 1, p. 5, 2015.

[P12] M. Staron, L. Kuzniarz, and C. Wohlin, "Empirical assessment of using stereotypes to improve comprehension of uml models: A set of experiments," *Journal of Systems and Software*, vol. 79, no. 5, pp. 727–742, 2006.

[P13] L. Muñoz, J.-N. Mazón, and J. Trujillo, "A family of experiments to validate measures for uml activity diagrams of etl processes in data warehouses," *Information and Software Technology*, vol. 52, no. 11, pp. 1188–1203, 2010.

[P14] F. Ricca, M. Di Penta, M. Torchiano, P. Tonella, and M. Ceccato, "How developers' experience and ability influence web application comprehension tasks supported by uml stereotypes: A series of four experiments," *IEEE Transactions on Software Engineering*, vol. 36, no. 1, pp. 96–118, 2010.

[P15] S. Mouchawrab, L. C. Briand, Y. Labiche, and M. Di Penta, "Assessing, comparing, and combining state machine-based testing and structural testing: a series of experiments," *IEEE Transactions on Software Engineering*, vol. 37, no. 2, pp. 161–187, 2011.

[P16] F. Ricca, G. Scanniello, M. Torchiano, G. Reggio, and E. Astesiano, "Assessing the effect of screen mockups on the comprehension of functional requirements," *ACM Transactions on Software Engineering and Methodology (TOSEM)*, vol. 24, no. 1, p. 1, 2014.

[P17] G. Scanniello, C. Gravino, M. Genero, J. Cruz-Lemus, and G. Tortora, "On the impact of uml analysis models on source-code comprehensibility and modifiability," *ACM Transactions on Software Engineering and Methodology (TOSEM)*, vol. 23, no. 2, p. 13, 2014.

[P18] J. Gonzalez-Huerta, E. Insfran, S. Abrahão, and G. Scanniello, "Validating a model-driven software architecture evaluation and improvement method: A family of experiments," *Information and Software Technology*, vol. 57, pp. 405–429, 2015.

[P19] G. Scanniello, C. Gravino, M. Risi, G. Tortora, and G. Dodero, "Documenting design-pattern instances: a family of experiments on source-code comprehensibility," *ACM Transactions on Software Engineering and Methodology (TOSEM)*, vol. 24, no. 3, p. 14, 2015.

[P20] A. M. Fernández-Sáez, M. Genero, D. Caivano, and M. R. Chaudron, "Does the level of detail of uml diagrams affect the maintainability of source code?: a family of experiments," *Empirical Software Engineering*, vol. 21, no. 1, pp. 212–259, 2016.

[P21] A. Porter and L. Votta, "Comparing detection methods for software requirements inspections: A replication using professional subjects," *Empirical software engineering*, vol. 3, no. 4, pp. 355–379, 1998.

[P22] O. Laitenberger, K. El Emam, and T. G. Harbich, "An internally replicated quasi-experimental comparison of checklist and perspective based reading of code documents," *IEEE Transactions on Software Engineering*, vol. 27, no. 5, pp. 387–421, 2001.

[P23] B. George and L. Williams, "A structured experiment of test-driven development," *Information and software Technology*, vol. 46, no. 5, pp. 337–342, 2004.

[P24] D. Pfahl, O. Laitenberger, G. Ruhe, J. Dorsch, and T. Krivobokova, "Evaluating the learning effectiveness of using simulations in software project management education: results from a twice replicated experiment," *Information and software technology*, vol. 46, no. 2, pp. 127–147, 2004.

[P25] L. Reynoso, M. Genero, M. Piattini, and E. Manso, "Assessing the impact of coupling on the understandability and modifiability of ocl expressions within uml/ocl combined models," in *11th IEEE International Software Metrics Symposium (METRICS'05)*. IEEE, 2005, pp. 10–pp.

[P26] F. Ricca, M. Di Penta, M. Torchiano, P. Tonella, M. Ceccato, and C. A. Visaggio, "Are fit tables really talking?" in *2008 ACM/IEEE 30th International Conference on Software Engineering*. IEEE, 2008, pp. 361–370.

[P27] C. G. Von Wangenheim, M. Thiry, and D. Kochanski, "Empirical evaluation of an educational game on software measurement," *Empirical Software Engineering*, vol. 14, no. 4, pp. 418–452, 2009.

[P28] J. A. Cruz-Lemus, M. Genero, D. Caivano, S. Abrahão, E. Insfrán, and J. A. Carsí, "Assessing the influence of stereotypes on the comprehension of uml sequence diagrams: A family of experiments," *Information and Software Technology*, vol. 53, no. 12, pp. 1391–1403, 2011.

[P29] M. Jørgensen, "Contrasting ideal and realistic conditions as a means to improve judgment-based software development effort

estimation," *Information and Software Technology*, vol. 53, no. 12, pp. 1382–1390, 2011.
[P30] M. A. Teruel, E. Navarro, V. López-Jaquero, F. Montero, J. Jaen, and P. González, "Analyzing the understandability of requirements engineering languages for cscw systems: A family of experiments," *Information and Software Technology*, vol. 54, no. 11, pp. 1215–1228, 2012.
[P31] P. Runeson, A. Stefik, A. Andrews, S. Gronblom, I. Porres, and S. Siebert, "A comparative analysis of three replicated experiments comparing inspection and unit testing," in *Replication in Empirical Software Engineering Research (RESER), 2011 Second International Workshop on*. IEEE, 2011, pp. 35–42.
[P32] T. Kosar, M. Mernik, and J. C. Carver, "Program comprehension of domain-specific and general-purpose languages: comparison using a family of experiments," *Empirical software engineering*, vol. 17, no. 3, pp. 276–304, 2012.
[P33] I. Hadar, I. Reinhartz-Berger, T. Kuflik, A. Perini, F. Ricca, and A. Susi, "Comparing the comprehensibility of requirements models expressed in use case and tropos: Results from a family of experiments," *Information and Software Technology*, vol. 55, no. 10, pp. 1823–1843, 2013.
[P34] C. R. L. Neto, I. do Carmo Machado, V. C. Garcia, and E. S. de Almeida, "Analyzing the effectiveness of a system testing tool for software product line engineering (s)," in *SEKE*, 2013.
[P35] A. Fernandez, S. Abrahão, and E. Insfran, "Empirical validation of a usability inspection method for model-driven web development," *Journal of Systems and Software*, vol. 86, no. 1, pp. 161–186, 2013.
[P36] P. Runeson, A. Stefik, and A. Andrews, "Variation factors in the design and analysis of replicated controlled experiments," *Empirical Software Engineering*, vol. 19, no. 6, pp. 1781–1808, 2014.
[P37] S. Ali, T. Yue, and I. Rubab, "Assessing the modeling of aspect state machines for testing from the perspective of modelers," in *2014 14th International Conference on Quality Software*. IEEE, 2014, pp. 234–239.
[P38] S. T. Acuña, M. N. Gómez, J. E. Hannay, N. Juristo, and D. Pfahl, "Are team personality and climate related to satisfaction and software quality? aggregating results from a twice replicated experiment," *Information and Software Technology*, vol. 57, pp. 141–156, 2015.
[P39] A. M. Fernández-Sáez, M. Genero, M. R. Chaudron, D. Caivano, and I. Ramos, "Are forward designed or reverse-engineered uml diagrams more helpful for code maintenance?: A family of experiments," *Information and Software Technology*, vol. 57, pp. 644–663, 2015.

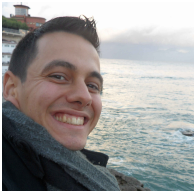
**Adrian Santos** received his MSc in Software and Systems and MSc in Software Project Management at the Technical University of Madrid, Spain, and his MSc in IT Auditing, Security and Government at the Autonomous University of Madrid, Spain. He is a PhD student at the University of Oulu, Finland. His research interests include empirical software engineering, agile methodologies, statistical analysis and data mining techniques. He is a member of the American Statistical Association (ASA) and the International Society for Bayesian Analysis (ISBA). For more information and details, please visit http://www.adriansantosparrilla.com

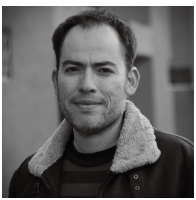
**Omar Gómez** received his BSc in Computer Engineering from the University of Guadalajara, his MSc in Software Engineering from the Center for Mathematical Research (CIMAT) and his PhD in Software and Systems from the Technical University of Madrid. He was with the University of Guadalajara (as adjunct assistant professor), the Autonomous University of Yucatan (as adjunct associate professor), the University of Oulu (as a postdoctoral research fellow) and the Technical School of Chimborazo (as a Prometeo-Senescyt researcher). He is currently an adjunct associate professor with the Technical School of Chimborazo. His main research interest is software engineering experimentation.

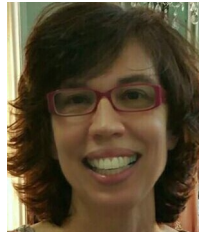
**Natalia Juristo** has been full professor of software engineering with the School of Computer Engineering at the Technical University of Madrid (UPM) since 1997. She was awarded a FiDiPro (Finland Distinguished Professor Program) professorship at the University of Oulu, starting in January 2013. She was the Director of the MSc in Software Engineering from 1992 to 2002 and coordinator of the Erasmus Mundus European Master on SE (with the participation of UPM, University of Bolzano, University of Kaiserslautern and Blege Institute of Technology) from 2006 to 2012. Her main research interests are experimental software engineering, requirements and testing. In 2001, she co-authored the book Basics of Software Engineering Experimentation (Kluwer). Natalia is a member of the editorial boards of IEEE Transactions on SE and Empirical SE, and Software: Testing, Verification and Reliability. She has served on several congress program committees (ICSE, RE, REFSQ, ESEM, ISESE, etc.), and has been congress program chair (EASE13, ISESE04 and SEKE97), as well as general chair (ESEM07, SNPD02 and SEKE01). Natalia was co-chair of ICSE Technical Briefings 2015 and co-chair of the Software Engineering in Practice (SEIP) track at ICSE 2017. She began her career as a developer at the European Space Agency (Rome) and the European Center for Nuclear Research (Geneva). She was a resident affiliate at the Software Engineering Institute in Pittsburgh in 1992. In 2009, Natalia was awarded an honorary doctorate by Blekinge Institute of Technology in Sweden. For more information and details, please visit http://grise.upm.es/miembros/natalia/